\documentclass[sigconf,nonacm]{acmart}

\usepackage{minted}
\usepackage{pifont} 
\usepackage{booktabs} 
\usepackage{tabularx}

\usepackage[table]{xcolor} 
\usepackage{graphicx} 
\usepackage{array} 
\usepackage[table,xcdraw]{xcolor} 
\usepackage{graphicx}
\usepackage{listings}
\usepackage{url}
\usepackage{subcaption} 
\usepackage{listings}
\lstdefinelanguage{Dockerfile}{
    morekeywords={FROM, RUN, CMD, COPY, WORKDIR, EXPOSE},
    sensitive=true,
    morecomment=[l]{\#},
    morestring=[b]"
}

\definecolor{darkgreen}{rgb}{0.0, 0.5, 0.0}

\usepackage{pgfplots}
\pgfplotsset{compat=1.18}

\usepackage{algpseudocode}    
\usepackage{float}  

\usepackage{color}     
\usepackage{caption}   

\definecolor{bggray}{rgb}{0.95,0.95,0.95} 

\setminted{
    fontsize=\footnotesize,
    linenos=true,
    frame=lines,
    framesep=2mm,
    bgcolor=bggray,
    breaklines=true,
    autogobble=true
}

\usepackage{array}
\usepackage{longtable}

\usepackage[ruled,vlined,linesnumbered]{algorithm2e}

\usepackage{listings}
\usepackage{tikz}
\usetikzlibrary{shapes.geometric, arrows.meta, positioning}

\usepackage{changepage}

\SetKwProg{Function}{Function}{}{end}
\SetNlSty{textbf}{\scriptsize}{}
\SetAlgoNlRelativeSize{-1}

\usepackage{tikz}
\usepackage{amsmath}

\usepackage{enumitem}
\setlist{noitemsep, topsep=0pt, leftmargin=*}

\AtBeginDocument{%
  }

\setcopyright{acmlicensed}
\copyrightyear{2025}
\acmYear{2025}
\acmDOI{XXXXXXX.XXXXXXX}

\acmISBN{978-1-4503-XXXX-X/18/06}

\usepackage{titlesec}

\titlespacing*{\section}
  {0pt}{2.5ex plus 1ex minus 0.5ex}{1.5ex plus 0.5ex}

\titlespacing*{\subsection}
  {0pt}{2.0ex plus 0.8ex minus 0.4ex}{1.2ex plus 0.4ex}

\titlespacing*{\subsubsection}
  {0pt}{1.8ex plus 0.6ex minus 0.3ex}{1.0ex plus 0.3ex}

\begin{document}

\title{Quantum Blockchain Survey: Foundations, Trends, and Gaps}

\author{Saurav Ghosh}
\affiliation{%
  \institution{Southeast Missouri State University}
  \city{Cape Girardeau}
  \state{Missouri}
  \country{USA}
}
\email{sghosh3s@semo.edu}

\author{Niloy Deb Roy Mishu}
\affiliation{%
  \institution{Indiana University Bloomington}
  \city{Bloomington}
  \state{Indiana}
  \country{USA}
}
\email{nilmish@iu.edu}

\begin{abstract}
Quantum computing poses fundamental risks to classical blockchain systems by undermining widely used cryptographic primitives. In response, two major research directions have emerged: post-quantum blockchains, which integrate quantum-resistant algorithms, and quantum blockchains, which leverage quantum properties such as entanglement and quantum key distribution. This survey reviews key developments in both areas, analyzing their cryptographic foundations, architectural designs, and implementation challenges. This work provides a comparative overview of technical proposals, highlight trade-offs in security, scalability, and deployment, and identify open research problems across hardware, consensus, and network design. The goal is to offer a structured and comprehensive reference for advancing secure blockchain systems in the quantum era.
\end{abstract}
\maketitle
\fancyhead{} 

\section{Introduction}
Quantum technologies are moving fast, and two of the most talked-about areas are the quantum internet and quantum blockchain. Both are still in early stages, but the ideas behind them could reshape how we think about secure communication, distributed systems, and trust. The quantum internet aims to connect quantum devices using entanglement and other quantum effects, while quantum blockchain tries to build tamper-proof ledgers using quantum principles or protect existing systems from quantum attacks.

There’s a lot of work happening in both areas, but much of it is still experimental or theoretical. At the same time, these fields are closely related: quantum networks could provide the foundation for secure blockchain communication, and blockchains could help organize distributed quantum systems. Still, there are many open questions. How do you build reliable routing for entanglement? Can we design consensus without classical trust assumptions? How do we layer security in a quantum-native stack?

In this review, a small set of papers will be focused, where each tackles a part of this broader problem. Some propose new blockchain models based on quantum effects, others look at quantum routing or layered security models. Each of them brings a different idea to the table. The goal is not just to explain what each paper says, but to look at how these ideas connect, what assumptions they make, and what gaps still exist.

\section{Backgrounds}

This section covers the key technical foundations needed to understand the systems discussed in this review. It introduces the core quantum primitives, the basic structure of quantum networks, and why integrating blockchain with quantum infrastructure has become an important research direction.

\subsection{Quantum Primitives}

Quantum entanglement is a key resource in both quantum communication and computation. It allows for correlations between particles that are stronger than anything classical systems can achieve. This property enables quantum key distribution (QKD), where two parties can share encryption keys with provable security based on quantum physics. QKD has been implemented over fiber and satellite links, but it still faces distance and throughput limitations.

While QKD protects against many types of attacks, it does not replace the need for digital signatures and consensus mechanisms. This is where post-quantum cryptography (PQC) comes in. PQC uses classical algorithms—like lattice-based or hash-based schemes—that are designed to resist attacks from quantum computers running algorithms such as Shor’s. These schemes are being standardized, but they often come with trade-offs in key size, signature length, or computational cost.

Entanglement and QKD are critical to how information is shared securely in a quantum network, while PQC focuses on protecting classical data from quantum attacks. These tools together shape how quantum and classical security models can work in parallel or be combined.

\subsection{Quantum Networks}

A quantum network is made up of nodes connected by links that distribute entangled qubits. These links can be optical fibers or satellite connections, and the network may include quantum repeaters to extend distances. Information isn't sent directly as data packets like in classical networks—instead, the goal is to create and maintain entanglement across the network, which can then be used for teleportation, secure key exchange, or distributed quantum computation.

The control plane of a quantum network coordinates entanglement generation, path selection, and resource management. This introduces unique challenges: entanglement is fragile, cannot be copied, and often decays quickly. As a result, routing and scheduling in quantum networks must deal with probabilistic link behavior and physical constraints that do not appear in classical systems.

Designing scalable quantum networks also means thinking in layers—similar to how the classical internet has protocol stacks. Efforts are underway to define such layered architectures, often drawing from software-defined networking (SDN) and classical internet design patterns, but adapted to quantum-specific requirements.

\subsection{Integration Drivers}

The idea of combining blockchain systems with the quantum internet is driven by the limitations of current distributed systems and the expected capabilities of future quantum infrastructure. Blockchains rely on consensus protocols and cryptographic primitives that are vulnerable to quantum attacks. At the same time, blockchains often assume trusted channels or delay-tolerant networks, which quantum communication could potentially improve.

On the other side, distributed quantum systems will eventually need mechanisms for coordination, auditability, and decentralized trust—goals that blockchain systems are already designed to support. Bringing these together opens up interesting design spaces: for example, using QKD to secure blockchain transactions, using quantum entanglement to prove timestamp integrity, or applying blockchain logic to organize entanglement routing and resource allocation.

This overlap has led researchers to explore hybrid systems, quantum-native blockchain models, and layered security architectures that treat blockchain and quantum networks as complementary technologies rather than separate domains.

\section{Thematic Analysis of Quantum Blockchain}
To better understand the research landscape in quantum blockchain, this section groups the key papers based on shared technical ideas and system designs. Instead of reviewing each work individually, we focus on the recurring themes that shape this field—such as the use of entanglement, quantum-secured consensus, hybrid architectures, and quantum routing. This thematic approach will make it easier to compare different directions and identify where future work is needed.

However, before going into the details of each work, it helps to have a quick view of what each paper focuses on and where its main contributions and limitations lie. Table~\ref{tab:survey_summary} summarizes the core set of papers covered in this review. This includes both foundational ideas and more recent proposals, spanning different parts of the broader quantum stack.

The table is organized to show each paper’s area of focus, the key ideas it introduces, and the challenges it leaves open. This overview is meant to provide context for the sections that follow, where each work is examined in more detail and discussed in relation to others.

\begin{table*}[htbp]
\centering
\normalsize
\renewcommand{\arraystretch}{1.5}
\setlength{\tabcolsep}{6pt}
\caption{Overview of Key Literature in Quantum Blockchain and Quantum Internet}
\label{tab:survey_summary}
\begin{tabular}{p{3cm} p{2.8cm} p{5.3cm} p{5.3cm}}
\toprule
\textbf{Reference} & \textbf{Area} & \textbf{Contributions} & \textbf{Limitations} \\
\midrule

Rajan \& Visser (2019) \cite{RajanVisser} & Quantum Blockchain & Introduced entanglement-in-time concept for quantum ledger integrity & Implementation remains theoretical; lacks scaling architecture \\

Kiktenko et al. (2018) \cite{Kiktenko} & Quantum Blockchain Security & Hybrid design using QKD and classical blockchain primitives & Requires quantum communication infrastructure not widely available \\

Sun et al. (2019) \cite{SunSopek} & Logic-based Quantum Blockchain & Integrated logic reasoning and quantum-enhanced signatures & No physical implementation; heavy reliance on untested consensus logic \\

Pant et al. (2019) \cite{Pant2019} & Quantum Routing & Developed model for entanglement-based routing over quantum networks & High dependency on link reliability; synchronous coordination needed \\

Yang et al. (2024) \cite{Yang2024} & Asynchronous Quantum Routing & Proposed DODAG-based routing using local state updates for entanglement paths & No real-world validation; assumes homogeneous node behavior \\

Shi \& Qian (2020) \cite{ShiQian2020} & Concurrent Entanglement Routing & Parallel entanglement routing design for high-throughput quantum networks & Link correlation and physical loss modeling not fully covered \\

Bernstein \& Lange (2017) \cite{BernsteinLange} & Post-Quantum Cryptography & Survey of lattice, hash-based, and code-based quantum-safe schemes & Trade-offs in key sizes and speed; unclear real-world performance \\

Xu et al. (2020) \cite{XuQKD} & QKD Systems & Comprehensive analysis of QKD protocol security and device imperfections & Distance limitations; low throughput hinders global adoption \\

Yang et al. (2023) \cite{YangSurvey} & Quantum Computing Survey & Structured overview of quantum hardware, networks, cryptography, and ML & Lacks empirical data or case studies; mainly theoretical \\

Lo, Curty \& Tamaki (2014) \cite{LoCurtyTamaki2014} & QKD Protocol Security & Formal security analysis of BB84 and decoy-based QKD implementations & Doesn't cover new QKD integration with quantum internet stacks \\

Tannu \& Qureshi (2019) \cite{TannuQureshi2019} & Quantum Hardware & Highlighted variability in qubit fidelity and performance-aware scheduling & Limited scalability; variability remains hard to mitigate in hardware \\
\bottomrule
\end{tabular}
\label{tab:survey_summary}
\end{table*}

\subsection{Foundations and Limitations}
Blockchains store records in a chain of blocks, where each new block references the hash of the previous one \cite{Nakamoto2008}. This design prevents tampering, as modifying one block requires changing all subsequent blocks. Classical blockchains rely on public-key cryptography and one-way hash functions. However, both are vulnerable to quantum algorithms—Shor’s algorithm can factor large numbers in polynomial time \cite{Shor}, and Grover’s algorithm offers quadratic speedups for brute-force search \cite{Grover1996}. A large enough quantum computer could forge signatures or find hash collisions easily \cite{Kearney}, which threatens blockchain integrity. These concerns highlight the need to explore solutions that can withstand quantum attacks.

A quantum blockchain uses quantum information to achieve ledger security \cite{RajanVisser}. This is not a simple upgrade. It involves methods such as quantum key distribution (QKD) \cite{Simon,Zhang,SunSopek} and entangled states to detect tampering and handle trust among distributed nodes. So far, these ideas exist mostly in theoretical research and small-scale tests \cite{RajanVisser,Edwards,Kiktenko,SunKulicki}, but they promise a way to secure blockchains against future quantum threats.

\subsubsection{Classical Cryptographic Vulnerabilities:}

Firstly, there is a public-key cryptography vulnerability. Systems like RSA or ECDSA assume that factoring large numbers is extremely hard. Shor’s algorithm can factor these in polynomial time \cite{Shor}, which weakens the security of digital signatures and may eventually allow attackers to sign transactions they do not own.

Secondly, there is the hash function weakening issue. Hashing prevents unauthorized changes to data because it is costly to find two inputs that produce the same output. Grover’s algorithm provides a quadratic speedup for brute-force attacks \cite{Grover1996}, weakening standard hash security and making brute-force collision finding much faster than before.

Thirdly, there is a consensus impact. Quantum computers may compute proof-of-work (PoW) puzzles more quickly \cite{SalmanEtAl2019,NoferEtAl2017}, which introduces vulnerabilities such as 51\% attacks (in PoW or PoS), quantum key theft (breaking RSA/ECDSA), and replay attacks. These are all possible to some extent under classical blockchain assumptions.

If the above concerns are not handled carefully, classical blockchains can no longer guarantee safe transactions in the future. An attacker with a powerful quantum computer could rewrite parts of the ledger, forge ownership, or compromise consensus protocols. A visual summary of this progression is shown in Figure~\ref{fig:blockchain_progression}.

\begin{figure*}[t]
    \centering
    \begin{tikzpicture}[
        node distance=0.5cm,
        every node/.style={align=center, font=\small, rectangle, draw, rounded corners, minimum width=4cm, minimum height=1.5cm},
        classical/.style={fill=blue!20, text=black},
        vulnerabilities/.style={fill=orange!40, text=black},
        quantum/.style={fill=purple!30, text=black},
        postquantum/.style={fill=teal!40, text=black},
        arrow/.style={thick, ->, >=latex, draw=black!70}
    ]

        \node[classical] (classical) {Classical Blockchain\\(e.g., RSA, ECDSA)};
        \node[vulnerabilities, right=of classical] (vulnerabilities) {Blockchain Vulnerabilities\\(Shor's, Grover's Algorithms)};
        \node[quantum, right=of vulnerabilities] (quantum) {Quantum Blockchain\\(QKD, Entangled States)};
        \node[postquantum, right=of quantum] (postquantum) {Post-Quantum Benefits\\(Quantum-Resilient, Tamper-Proof)};

        \draw[arrow] (classical) -- (vulnerabilities);
        \draw[arrow] (vulnerabilities) -- (quantum);
        \draw[arrow] (quantum) -- (postquantum);
    \end{tikzpicture}
    \caption{A linear progression showing the evolution from classical blockchain to post-quantum benefits.}
    \label{fig:blockchain_progression}
\end{figure*}

\subsubsection{Key Components:}

Although quantum blockchain designs vary, most share the following components:

\textbf{Quantum-Resistant Cryptography:} Since traditional methods may be broken by quantum computers, post-quantum cryptography—such as lattice-based or error-correcting code-based approaches—is proposed \cite{BernsteinLange,FernandezCarames}. Another strategy is to use fully quantum approaches like quantum keys and one-way transformations that are impossible to reverse without a trapdoor \cite{RajanVisser,Edwards,Kiktenko,SunKulicki}.

\textbf{Entangled State Storage:} Some designs propose linking blocks through entangled quantum states such as GHZ states. This means that if a block is entangled with the previous one, any tampering would collapse the entanglement and be instantly detectable \cite{Greenberger,Carvacho,Megidish}.

\textbf{New Consensus Mechanisms:} In a quantum blockchain, consensus needs to be reimagined. Instead of mining or staking, some systems use quantum random number generators \cite{McCutcheon} to ensure true unpredictability in block selection. Others propose interactive verification tests, where multiple nodes validate quantum-generated blocks collaboratively \cite{RajanVisser,Edwards,Kiktenko,SunKulicki}.

\textbf{Quantum Network Challenges:} Running a quantum blockchain is not just a software problem—it requires specialized hardware. Real-world quantum networks depend on technologies such as photon-based repeaters or superconducting qubits. However, quantum states are fragile; noise and decoherence can degrade the quality of stored and transmitted data \cite{Simon,Zhang,SunSopek}.

\subsubsection{Implementation Challenges:}

Shor’s and Grover’s algorithms remain the most prominent threats to classical security \cite{Shor,Grover1996}. While some mitigation can be achieved by increasing key and hash lengths, this is unlikely to provide a long-term solution once scalable quantum hardware becomes available \cite{Kearney}.

Current quantum processors are still early in development, often with only dozens or hundreds of qubits. IBM, for example, has systems with 127 qubits. These processors suffer from high error rates, limiting their practical use \cite{TannuQureshi2019,RodenburgPappas2017}. Nonetheless, hardware is improving, and the goal of building thousands or millions of stable qubits remains a central research focus.

Quantum states themselves are fragile and collapse when observed. Entangled states can lose consistency over time \cite{Bruschi,Banerjee,Nowakowski,RajanArxiv}, which presents a serious challenge for blockchain designs that rely on entanglement. The verification process must avoid unintentionally destroying the very state being validated. Moreover, sustaining large-scale entanglement demands sophisticated quantum engineering.

Consensus over quantum channels introduces further complexity. Multi-party protocols may be needed, where nodes measure different parts of a quantum state with randomly chosen bases. The measurement results must align with expected patterns to confirm validity. This makes consensus more complex than in classical systems and raises concerns about performance and scalability.

\subsubsection{Critical Observations:}

Quantum computing fundamentally shifts the assumptions behind blockchain security, but implementing a full quantum blockchain requires advanced physics. This includes stabilizing qubits and deploying quantum networks that can tolerate distance and noise. Many projects do not yet have the resources or expertise to realize such systems. Meanwhile, the classical blockchain community is well-established, whereas quantum blockchain development is still early-stage. Researchers must build prototypes and refine protocols without sacrificing speed, which will take time and significant investment.

Some research papers inaccurately describe Shor’s (1994) and Grover’s (1996) algorithms as “recent advances,” despite their age. The true concern is the increasing pace of hardware progress. Moreover, several papers overlook critical issues like state stability and scalable quantum consensus. Quantum blockchains require enormous computational resources and coordination, which can limit adoption even among large organizations. A clearer treatment of these concerns would provide a more balanced view.

Quantum computers will almost certainly compromise traditional blockchain security. To address this, post-quantum cryptography and quantum-native designs must be actively explored. Both offer promise, but also face significant engineering and design challenges. Entangled ledgers, quantum-proof signatures, and reimagined consensus mechanisms are theoretically compelling, but remain unproven at scale. For now, research continues at the intersection of theoretical innovation and small experimental systems. By preparing early, the field aims to avoid a disruptive crisis when quantum attacks become viable.

\subsection{Computing \& Networking Challenges}

A broad overview of quantum computing and communications from a computer science perspective \cite{YangSurvey} is essential to understanding the field. Therefore, it is important to explore four key areas that are expected to form the future of the quantum Internet: quantum computers, quantum networks, quantum cryptography, and quantum machine learning. This section discusses their historical development and addresses major questions related to feasibility, performance, and security. Table~\ref{tab:quantum_subfields} summarizes the core technical challenges and current status across these subfields, highlighting their interdependence in realizing a scalable and secure quantum infrastructure.

{\normalsize
\begin{table*}[h]
    \centering
    \renewcommand{\arraystretch}{1.4}
    \setlength{\tabcolsep}{8pt}
    \caption{Core Technical Challenges Across Quantum Subfields}
    \label{tab:quantum_subfields}
    \begin{tabularx}{\textwidth}{l X X}
        \toprule
        \textbf{Subfield} & \textbf{Key Technical Challenge} & \textbf{Current Status} \\
        \midrule
        Quantum Hardware & High-fidelity multi-qubit operations, scalability, cryogenic control integration & Advancing via superconducting and trapped-ion platforms (IBM, Google, IonQ) \\
        Quantum Networking & Stable long-distance entanglement distribution, quantum repeater deployment & Prototype networks under test; early QKD deployments in metropolitan areas \\
        Quantum Cryptography & Efficient quantum key distribution (QKD), integration with classical infrastructure & Actively standardized (e.g., ETSI, ITU); limited real-world adoption \\
        Quantum Machine Learning & Variational circuit stability, hybrid algorithm tuning on NISQ devices & Early-stage with limited scalability; constrained by hardware noise and data loading \\
        \bottomrule
    \end{tabularx}
\end{table*}
}

\subsubsection{Complexity of Quantum Research:}

Quantum technologies have made rapid progress, with universal quantum computers now supporting hundreds of qubits and quantum annealers reaching thousands \cite{Chow,Roberts}. Simultaneously, advances in quantum networking, cryptography, and machine learning have made the field more diverse. Without a structured guide, researchers and developers can easily become overwhelmed.

This work explains how quantum mechanics allows computation through principles like superposition and entanglement. It identifies key problems in building scalable and fault-tolerant quantum hardware \cite{Reilly}, describes how quantum networks use entanglement, quantum repeaters, and specialized routing to connect quantum devices across distances \cite{Pant2019}, reviews quantum cryptographic techniques such as quantum key distribution (QKD) \cite{XuQKD}, and explores quantum machine learning applications, including how quantum computing can speed up optimization problems or improve data analysis through hybrid quantum-classical methods \cite{PerdomoOrtiz}.

\subsubsection{Methodological Foundations:}

Quantum computers offer exponential speed-ups through principles like superposition and entanglement. However, noise and decoherence disrupt quantum states, and the number of available qubits remains too small for large-scale computations \cite{TannuQureshi2019}. Error correction is also difficult, as maintaining fault tolerance while scaling up quantum processors poses a significant engineering challenge.

Quantum networks rely on entanglement distribution via Einstein-Podolsky-Rosen (EPR) pairs and quantum repeaters \cite{Sangouard}. Connecting quantum devices over long distances is complex due to fragile quantum states, signal loss, decoherence, and the challenge of building efficient repeaters \cite{Pant2019}. Unlike classical signals, quantum information cannot be copied or amplified, necessitating new routing methods for probabilistic entanglement and teleportation.

\subsubsection{Quantum Cryptography:}
Quantum computing threatens current cryptographic schemes like RSA and ECDSA, making quantum-safe alternatives essential.

\textbf{Post-quantum cryptography:} It relies on lattice-based or code-based techniques \cite{BernsteinLange,FernandezCarames}.

\textbf{Quantum key distribution (QKD):} It ensures secure communication using quantum principles, as demonstrated in BB84 and E91 protocols \cite{BennettBrassard,Ekert1991}.

Other approaches, such as quantum signatures and coin-flipping, remain theoretical but highlight broader potential \cite{BroadbentSchaffner}.

However, quantum cryptography has drawbacks. Despite theoretical unbreakability, real-world QKD systems face low key generation rates, DoS attack susceptibility, and operational distance limitations \cite{XuQKD}, all of which challenge its global practicality.

\subsubsection{Quantum Machine Learning:}
Quantum hardware may accelerate tasks like large-scale optimization using quantum annealing and variational circuits. Quantum machine learning aims to enhance data analysis, while classical machine learning can optimize quantum experiments. However, today’s quantum processors are too noisy and lack sufficient qubit connectivity for large, real-world datasets. Designing scalable, hybrid algorithms remains an open problem.

\subsubsection{Integration Challenges:}

Integrating quantum computing, networking, cryptography, and machine learning into a unified system is a major challenge. Each subfield has different performance requirements, noise sensitivities, and hardware constraints. Ensuring system compatibility demands overcoming fundamental issues and developing standardized protocols.

While each subfield offers breakthroughs, the path to functional quantum networking remains difficult due to scalability, stability, and integration challenges. Figure \ref{fig:fishbone-integration} illustrates the integration issues briefly.

\begin{figure}[h]
    \centering
    \resizebox{\columnwidth}{!}{
        \begin{tikzpicture}[
            >=latex,
            node distance=2cm and 2cm,
            every node/.style={font=\sffamily},
            fishbone/.style={thick},
            cause/.style={thin, -{Triangle[open,length=4pt]}},
            branch/.style={anchor=east, text width=3cm, align=left, font=\small}
        ]

        \draw[fishbone] (-5.5,0) -- (3,0);

        \draw[cause] (-5,0) -- ++(1.5,1) node[branch, above=0.3cm] {Diverse \\ Performance \\ Requirements};
        \draw[cause] (-3,0) -- ++(1.5,1) node[branch, above=0.3cm] {Noise \\ Sensitivity};
        \draw[cause] (-1,0) -- ++(1.5,1) node[branch, above=0.3cm] {Scalability \\ Issues};

        \draw[cause] (-5,0) -- ++(1.5,-1) node[branch, below=0.3cm] {Hardware \\ Compatibility};
        \draw[cause] (-3,0) -- ++(1.5,-1) node[branch, below=0.3cm] {Standardization \\ Gaps};
        \draw[cause] (-1,0) -- ++(1.5,-1) node[branch, below=0.3cm] {High Resource \\ Overhead};

        \node[draw, rectangle, minimum width=3cm, minimum height=1cm, fill=gray!20] at (3,0) {Integration Challenges};

        \end{tikzpicture}
    }
    \caption{Challenges in Integrating Quantum and Classical Systems}
    \label{fig:fishbone-integration}
\end{figure}

\subsubsection{Open Research Questions:}

\textbf{Quantum Hardware Constraints:}
Quantum computers still have limited stable qubits. Noise and decoherence shorten coherence times, and low-temperature operation is required \cite{TannuQureshi2019}. Improved hardware is necessary to move toward practical systems.

\textbf{Networking Complexity:}
Creating a global quantum Internet is difficult. Entanglement weakens over distance, and quantum routers must re-establish links continuously. No-cloning prohibits signal amplification, limiting scalability \cite{ChenSpace}.

\textbf{Security Versus Performance Tradeoff:}
Quantum computing threatens current cryptographic systems. While post-quantum methods exist, they may reduce performance and demand infrastructure overhauls \cite{Mosca}.

\textbf{Quantum Machine Learning Limitations: }Quantum machine learning is promising but limited by current hardware and dataset compatibility. More research is required to realize practical applications \cite{Biamonte}.

Progress depends on strong theory, significant investment, and standards for integration. Quantum systems must be combined with classical infrastructure, which adds complexity beyond scientific research \cite{Reilly}.

These works however lacks implementation detail. They does not fully explain quantum-classical interaction, especially error correction and hardware challenges. They also lacks a roadmap for scaling from prototype to large-scale systems.

A few real-world case studies are discussed as well. Although theoretical models are explored, practical examples or benchmarks are missing. Security discussions largely focus on well-known quantum attacks, omitting side-channel vulnerabilities and the deployment of post-quantum cryptography at scale.

The future of quantum computing depends on solving these limitations—advancing hardware, securing networks, and integrating with classical systems. Industry interest is growing, particularly in areas like quantum cloud services and QKD. With improvements in error correction, entanglement distribution, and post-quantum cryptography, a global quantum Internet may emerge within the next decade.


\subsection{Post-Quantum vs. Quantum Blockchain}

The main challenge facing blockchain technology is the rapid progress in quantum computing. In particular, the problems inherent in our current cryptographic and consensus systems need to be discussed before outlining how post-quantum and quantum blockchains can address these issues.

\subsubsection{Problem Identification:}

Classical blockchains depend on public-key cryptography (such as RSA and ECDSA) and robust hash functions (like SHA-256) to secure transactions and maintain system integrity \cite{Nakamoto2008,Johnson2001}. However, quantum algorithms—most notably Shor’s algorithm \cite{Shor} for factoring and Grover’s algorithm \cite{Grover1996} for search—pose serious risks. Quantum computers can break these cryptographic methods. If an adversary uses Shor’s algorithm to factor large integers, they could derive private keys from public keys, thereby compromising the entire blockchain system.

\textbf{Consensus Vulnerabilities:}
Blockchain networks rely on consensus mechanisms, such as proof-of-work, to validate transactions and maintain a decentralized ledger \cite{Nakamoto2008}. Quantum computing may allow an attacker to solve proof-of-work puzzles much faster than classical computers, potentially leading to a 51\% attack. This disparity could let a malicious actor control the network, rewrite history, or disrupt the consensus process.

\textbf{Hardware and Networking Limits:}
Current quantum hardware is still in its infancy—limited qubit counts, high noise levels, and rapid decoherence hamper practical large-scale computations. Additionally, while quantum key distribution (QKD) offers a secure means for exchanging keys, distributing entangled states over long distances is extremely challenging due to signal loss and the no-cloning theorem \cite{Bennett2014,Ekert1991}. Without scalable and stable quantum hardware and networking, fully quantum blockchain systems remain largely theoretical.

\textbf{Performance Trade-offs:}
Replacing classical cryptography with post-quantum alternatives introduces performance challenges. Many post-quantum schemes require larger key sizes and more computation, which can slow down transaction processing and complicate integration with existing infrastructure.

\subsubsection{Solution Approaches:}

After understanding the problems, the two main approaches to secure blockchain systems in a quantum era are outlined below.

\paragraph{Post-Quantum Blockchains:}
\begin{itemize}
    \item \textbf{What:} Replace vulnerable classical cryptographic primitives with quantum-resistant alternatives.
    \item \textbf{Why:} These rely on mathematical problems that are hard for quantum computers (e.g., lattice-based or code-based schemes) \cite{BernsteinLange}.
    \item \textbf{How:} For instance, lattice-based schemes based on the Shortest Vector Problem (SVP) or Learning With Errors (LWE) \cite{Ajtai1996,Regev2009} are proposed. They secure transactions but increase computational overhead.
\end{itemize}

\paragraph{Quantum Blockchains:}
\begin{itemize}
    \item \textbf{What:} Rebuild blockchain architecture using quantum technologies.
    \item \textbf{Why:} Employ QKD, quantum digital signatures, and quantum data structures for security grounded in quantum mechanics \cite{Kiktenko2018, Rajan2019}.
    \item \textbf{How:} 
    \begin{itemize}
        \item \textbf{Hybrid Quantum Blockchains:} Combine classical infrastructure with quantum techniques (e.g., QKD) \cite{JPMorgan2018}.
        \item \textbf{Fully Quantum Blockchains:} Represent each block as a quantum state linked by entanglement \cite{Rajan2019}. Practical deployment remains constrained by current hardware.
    \end{itemize}
\end{itemize}

\begin{table*}[h]
    \centering
    \renewcommand{\arraystretch}{1.8}
    \setlength{\tabcolsep}{12pt}
    \rowcolors{2}{gray!10}{white}
    \begin{tabular}{p{2.8cm} p{3.6cm} p{3.8cm} p{4.7cm}}
        \toprule
        \rowcolor{blue!20}
        \textbf{Characteristic} & \textbf{Classical Blockchain} & \textbf{Quantum Blockchain} & \textbf{Post-Quantum Blockchain} \\
        \midrule
        \textbf{Security} & Vulnerable to quantum attacks (Shor, Grover) & Uses quantum cryptography (QKD, entanglement) & Resistant to quantum attacks (Lattice, Hash-based cryptography) \\
        \textbf{Consensus} & Proof-of-Work / Proof-Of-Stake & Quantum-secured consensus (Quantum RNG, QKD) & Modified PoW/PoS with post-quantum cryptographic methods \\
        \textbf{Performance} & High throughput in classical systems & Experimental, requires quantum resources & Higher computational cost due to larger key sizes \\
        \textbf{Scalability} & Well-tested, mature infrastructure & Limited by quantum hardware challenges & Still emerging, performance improvements ongoing \\
        \textbf{Implementation Readiness} & Fully deployed in real-world applications & Mostly theoretical, few small-scale experiments & Actively researched, partial real-world deployment \\
        \bottomrule
    \end{tabular}
    \caption{Comparison of Blockchain Approaches: Classical, Quantum, and Post-Quantum}
    \label{tab:blockchain_comparison}
\end{table*}

\subsubsection{Issue Significance:}

These problems strike at the core of blockchain trust:
\begin{itemize}
  \item If cryptographic methods fail, the trust model collapses.
  \item A compromised consensus mechanism undermines decentralization.
  \item Limitations in current quantum technology hinder practical deployment.
  \item Performance trade-offs discourage the adoption of new cryptographic methods.
\end{itemize}

A clear understanding of these issues is critical. Whether upgrading classical systems with post-quantum cryptography or developing quantum-native blockchains, addressing these problems is key to securing decentralized systems in the quantum era.


\subsection{Quantum Internet Routing Challenges}

As we know, the future of the Computing Industry will be in Quantum Computing, and as a by-product, a lot of work is already going on in the Quantum Internet sphere \citep{Kimble2008,Wehner2018,GisinThew2007,Komar2014,Yang2023}. However, for efficiency on the Internet, we need advanced quantum networking techniques that have significantly risen. Routing quantum entanglement without relying on fixed time slots is a big challenge because quantum links are really unpredictable and short-lived \citep{Pant2019,Chakraborty2019}. Traditional methods that use synchronized time slots can waste resources because entanglement doesn't always happen at the right time. Therefore it is important to explore asynchronous routing, which means that the network updates its connections in real-time rather than following a fixed schedule. Quantum repeaters play an important role here, as they help extend entanglement over long distances \citep{Briegel1998,Pan1998}. However, these repeaters are fragile as well, and entanglement swapping is not always successful. The best way to connect two distant nodes is to let each part of the network make local decisions as soon as entanglement becomes available. I felt this can justify a better use of quantum resources and therefore improve the chances of successfully routing entanglement over long distances. Finally, the most important more here is: What is the best way to connect two remote nodes when each quantum link is probabilistic and has a limited lifetime? 

\subsubsection{Asynchronous Quantum Routing Complexity:}

The problem here is very practical. Quantum networks need to create connections over long distances using repeaters. Traditional methods use synchronized time slots, where all nodes try to create and swap entanglement at the same time. But this wastes many entangled pairs because not all attempts succeed. Since quantum states disappear quickly \citep{DegenReinhardCappellaro2017}, wasted entanglement makes the network inefficient. A better way would be an asynchronous approach, where each node updates its connections independently based on local information. This is like how classical networks find paths using distributed graphs. With this method, quantum networks can save resources and improve their chances of creating stable connections over long distances.

\subsubsection{Asynchronous Routing Techniques:}

This work has been built on three main components, as illustrated in Figure \ref{fig:async_routing}. First, it uses quantum repeaters which generate entangled pairs over direct links between adjacent nodes. These repeaters do entanglement swapping so that a connection between distant nodes can be built from many short links \citep{AzumaTamakiLo2015}. Second, it introduces a system that keeps an ongoing update of these connections in a network graph, which can be a destination-oriented directed acyclic graph (DODAG) or a spanning tree. Instead of waiting for all nodes to update together, each node independently updates its status based on what it sees around it \citep{Beals2013,VanMeterDevitt2016}. Third, the approach mixes quantum operations (like creating and swapping entanglements) with classical operations (i.e., sending out routing information). Nodes listen for connection requests and use their local network information to choose the next node to connect with. This method saves unused entanglements for future use and avoids making all nodes act at the same time.

This approach is built on many well-known ideas as well. For instance, the concept of quantum-native repeaters was introduced by Briegel \textit{et al.} \citep{Briegel1998} and further studied in some later works \citep{Pan1998,Tchebotareva2019}. The idea of using distributed graph structures for routing is taken from classical networking research \citep{Alexander2012,Gallager1983}.

\begin{figure*}[h]
    \centering
    \fbox{
        \begin{minipage}{0.85\linewidth}
            \centering
            \textbf{\textcolor{blue}{Asynchronous Entanglement Routing Components}} \\[0.3cm]
            \renewcommand{\arraystretch}{1.5} 
            \begin{tabular}{|c|}
                \hline
                \rowcolor{gray!20} \textbf{\textcolor{black}{Quantum Repeaters: Generate entangled pairs over direct links \citep{AzumaTamakiLo2015}.}} \\ 
                \hline
                \rowcolor{cyan!20} \textbf{\textcolor{black}{Network Graph Update: Nodes update DODAG or spanning tree asynchronously \citep{Beals2013,VanMeterDevitt2016}.}} \\ 
                \hline
                \rowcolor{blue!20} \textbf{\textcolor{black}{Hybrid Quantum-Classical Operations: Mix entanglement swapping with classical routing data.}} \\ 
                \hline
            \end{tabular}
        \end{minipage}
    }
    \caption{Key Components of Asynchronous Entanglement Routing}
    \label{fig:async_routing}
\end{figure*}

\subsubsection{Components and Challenges:}

Each node in the network tries to create a quantum link with its neighbors, succeeding with a chance indicated by the parameter $p$. Swapping connections at intermediate nodes, needed to extend the link further, also has a success rate, represented by $q$. Nodes only know what's happening with their direct connections, which complicates finding the best path across the whole network. They solve this by updating a shared map of connections on their own, using only local data. Whenever a node needs to extend the connection, it uses this map to decide where to send the request next.

The big challenges here are the unpredictability of quantum operations because, as we know, the connections can fail to establish or drop unexpectedly and the fact that quantum links don't last long. Also, delays in updating the map can cause the route decisions to be based on old, incorrect information. Lastly, since each node updates its own map, the system has to be smart about avoiding loops and dead ends. The design takes ideas from classical distributed routing algorithms \citep{Pant2019,Chakraborty2019} to avoid these issues.

The asynchronous approach can significantly improve the entanglement rate compared to synchronous methods. In synchronous protocols, all nodes are forced to use up available entanglements at fixed times, which can lower the overall rate. With asynchronous routing, unused direct-link entanglements are preserved and can be reused in future routing attempts. This results in a higher upper bound for the entanglement rate. The improvement is even more noticeable when the coherence time increases. It is needless to say as soon as we see the quantum hardware advances and coherence times become longer \citep{VanMeterDevitt2016}, asynchronous routing protocols may become essential.

There are still open challenges, however. For example, this work assumes that all nodes and links are homogeneous. In real networks, link qualities and node performances vary widely. Also, while they showed simulation results on grid topologies, the performance on more irregular networks is not fully explored. Finally, they did not have a good discussion on the security aspects of the protocol. In a practical network, delays and asynchronous updates might open new vulnerabilities that might need some additional attention.

Apart from the open challenges, these work does not go into detail about how the classical update process integrates with the quantum operations. The delay in classical communication is a critical factor, and, given the short coherence times, it is much more difficult in the event of quantum operations; the protocol does not fully address how to mitigate this issue. Security concerns, for example, the risk of malicious behavior in an asynchronous environment, are not analyzed as well.

The introduction of an asynchronous entanglement routing protocol for quantum networks works by updating a shared network graph using local information, making it both simple and effective. It outperforms traditional synchronous methods by saving entanglement resources and increasing the entanglement rate. However, further testing in real-world conditions is needed to account for differences in network components and potential security risks. Overall, this study is an important step toward a scalable and efficient Quantum Internet.

However, these works are important because by proposing an asynchronous entanglement routing protocol, they made a great contribution to the field of quantum networking. The idea of maintaining a distributed graph with local information is both simple and effective. This method shows a clear advantage over traditional synchronous approaches. However, it is also important to test the protocol under realistic conditions. In one line, this work can be considered a great step towards an efficient Quantum Internet.

\subsection{Entanglement Distribution}

Given that the future is in Quantum Computing, we need to improve end-to-end entanglement in large quantum networks. This paper has used a multi-tree routing method that builds several destination-oriented directed acyclic graphs (DODAGs) at the same time. This method works for different network layouts i.e., grids, barbells, and realistic networks such as ESnet and Internet2. It is needless to say, by using multiple trees instead of a single one, it is very much possible to get a better end-to-end entanglement rate. Based on the simulations illustrated, it can be seen that the multi-tree method gives higher entanglement rates than both the single-tree method and the traditional synchronous methods \citep{Wehner2018,Yang2023}.

\subsubsection{Multi-Tree Quantum Routing Complexity:}

The main problem in entanglement distribution is to create end-to-end entanglement over long distances. In a quantum network, nodes must swap entanglements using intermediate repeaters \citep{Briegel1998,Pan1998}. In many methods, the routing is divided into two phases. In the external phase, nodes create direct entanglements with each other. In the internal phase, repeaters swap these entanglements to connect distant nodes \citep{Pant2019,ShiQian2020}. However, using fixed time slots in these two phases can waste entanglement because the quantum links are not stable and the results are unpredictable. It can easily be noticed however that, in an asynchronous method, where each node updates its own view in real-time, can make better use of the available entanglement.

\subsubsection{Background of Multi-Tree Routing:}

This work has been built on their earlier asynchronous routing protocol \citep{Yang2024}. They started by forming a tree-like structure from the available direct links. This tree may be a DODAG or a spanning tree. Each node uses only the local information it gathers from its neighbors, much like classical routing methods \citep{Alexander2012,Gallager1983}. Then, instead of using one tree, they formed several trees (a multi-tree or DODAG forest). With multiple trees, nodes in dense areas can choose the best path. This structure helps avoid long detours that occur when using a single central root. In their method, quantum operations (for entanglement generation and swapping) work together with classical operations (for exchanging routing messages) to keep the network graph updated\citep{FurnasZacks1994,ErdsRnyi1959}.

In these works, however, every node here tries to create a direct link with its neighbors. As noted alike the previous one, the chance of success for each direct link is given by $p$, and the chance of success for entanglement swapping at a repeater is given by $q$. Each node only sees the links to its immediate neighbors. They then update a local map of the network, called the instant topology. When a node needs to forward a connection request, it looks at this map to choose the best next node. The main challenge is that quantum links are probabilistic and decay quickly. If the local map is not updated fast enough, the routing decision may use old information. Also, when many trees share the same area, loops, and redundant paths might occur. Our design uses simple rules, such as each node choosing only one parent per tree, to avoid these problems \citep{Pant2019,ShiQian2020}.

It is also observed that the multi-tree approach gives better results than the single-tree method or the traditional synchronous routing. In synchronous protocols, all nodes swap entanglement at fixed times. This fixed schedule can use up all the available entanglements even when some links are not ready, which lowers the overall rate \citep{Pant2019}. With an asynchronous multi-tree method, unused direct-link entanglements are saved for later use. This saves resources and raises the maximum achievable entanglement rate. The benefit becomes even clearer when the coherence time of the quantum links is longer \citep{VanMeterDevitt2016}. 

As noted in the previous challenge, these works assume that all links and nodes are the same, but in real networks, they vary. Also, the simulations mainly use grid and barbell topologies. More tests are needed on irregular networks.

It is true that, through this multi-tree quantum routing, a promising way to improve end-to-end entanglement rates in large networks is established. The idea of using multiple DODAGs to form a distributed graph with local updates helps save entanglement and raises the overall rate. Further research that will deal with network heterogeneity and address the security concerns is needed, however. 
\subsection{Quantum Internet Security Models}

As we know, the quantum internet holds great promise for secure communication and advanced computing. However, it also faces unique security challenges that must be solved to protect its Confidentiality, Integrity, and Availability (CIA). Features like superposition bring new types of vulnerabilities that do not exist in classical networks. Related previous papers did not discuss security issues; however, this paper looks at the security challenges in the different layers of the quantum internet: the physical, link, network, and application layers. They have, however, used only the vulnerabilities and mitigation techniques that have been reported in the literature to build a framework that mixes classical and quantum methods for protecting the network \citep{Yang2023,LoCurtyTamaki2014}.

\subsubsection{Security Challenges in Internet:}

Quantum mechanics gives us benefits such as the no-cloning theorem and measurement-induced disturbance, which help detect eavesdropping \citep{Ekert1991}. On the other side, these same features create new challenges. For example, while Quantum Key Distribution (QKD) can offer unconditional security, it can still be attacked by methods such as Photon Number Splitting (PNS) and Trojan-horse attacks \citep{Brassard2000,Zhao2008}. This work has found out that even though quantum protocols promise very high security, the hardware is often noisy and the classical channels used for control and routing may add extra weaknesses. This layered nature of the quantum internet makes it necessary to study security risks at every level.


The approach in these studies was to break down the quantum internet into layers, much like the classical network protocol stack. They have identified the following layers:
\begin{itemize}
    \item The \textbf{physical layer} > Quantum memories and the transmission media.
    \item The \textbf{link layer} > Handles the direct transmission of qubits and the operations of quantum repeaters.
    \item The \textbf{network layer} > Deals with the routing of entanglement.
    \item The \textbf{application layer} > Supports quantum applications such as QKD, quantum teleportation, and distributed quantum computing \citep{Illiano2022}.
\end{itemize}
It is very important to discuss the main components at each layer, as illustrated in Figure \ref{fig:security_framework} and the specific attacks that target them. It is also very important to review the mitigation techniques that others have proposed. 

\begin{figure}[h]
    \centering
    \begin{tikzpicture}
        \node (app) [draw, fill=blue!20, minimum width=6cm, minimum height=1cm] {Application Layer Security};
        \node (net) [draw, fill=green!20, minimum width=6cm, minimum height=1cm, below=0.3cm of app] {Network Layer Security};
        \node (link) [draw, fill=orange!20, minimum width=6cm, minimum height=1cm, below=0.3cm of net] {Link Layer Security};
        \node (phy) [draw, fill=red!20, minimum width=6cm, minimum height=1cm, below=0.3cm of link] {Physical Layer Security};
    \end{tikzpicture}
    \caption{Layer-Wise Security Framework for the Quantum Internet}
    \label{fig:security_framework}
\end{figure}

\subsubsection{Components and Challenge:}

These work breaks the security framework into 4 main layers. In the physical layer, the main challenges come from the fragile nature of quantum devices. Photon-based systems can be attacked by PNS or Trojan-horse methods \citep{Zhao2008,Lu2022}. These attacks try to steal information by abusing imperfections in photon sources and detectors. 

In the link layer, the focus is on the direct channels between nodes. Quantum repeaters are used to extend the range of entanglement. However, these repeaters can be attacked by methods such as entangling-probe attacks or even man-in-the-middle attacks if an attacker takes control of a node \citep{Briegel1998,Pan1998}. 

In the network layer, the challenge is to route entanglement over complex network layouts. Many routing schemes use classical control that is synchronized across the network, which creates single points of failure. If, for example, an attacker can compromise a repeater or change routing messages, the end-to-end entanglement can be seriously affected \citep{Pant2019,ShiQian2020}. Finally, at the application layer, protocols like BB84 and entanglement-based QKD are vulnerable to advanced attacks. Even though these protocols have ways to detect eavesdropping, they still face side-channel attacks and timing errors \citep{Ekert1991,LoCurtyTamaki2014}. A brief version is illustrated in Figure \ref{tab:security_threats}. 

\begin{table*}[h]
    \centering
    \large
    \renewcommand{\arraystretch}{1.5}
    \caption{Security Threats and Mitigation Strategies Across Quantum Internet Layers}
    \label{tab:security_threats}
    \begin{tabularx}{\textwidth}{l X X}
        \toprule
        \textbf{Layer} & \textbf{Security Risks} & \textbf{Mitigation Strategies} \\
        \midrule
        \textbf{Physical} & 
        Qubit decoherence, thermal noise, photon loss, side-channel leakage from quantum hardware (e.g., emission timing, power traces) & 
        Quantum error correction, low-temperature shielding, fault-tolerant design, side-channel-resistant hardware interfaces \\
        
        \textbf{Link} & 
        Interception of quantum states (quantum man-in-the-middle), channel tampering during entanglement distribution, measurement disturbance & 
        Quantum authentication protocols, decoy-state QKD, entanglement purification, Bell-test-based link verification \\
        
        \textbf{Network} & 
        Malicious or compromised nodes disrupting entanglement routing, denial-of-service through entanglement flooding, manipulation of routing metadata & 
        Quantum-aware routing (e.g., entanglement-aware Dijkstra), authenticated control-plane messages, multi-path redundancy, topology-aware monitoring \\
        
        \textbf{Application} & 
        Trojan-horse attacks (e.g., injecting light to extract info), delay/timing-based side-channel leakage, protocol misuse, impersonation & 
        Device-independent quantum cryptography, quantum-safe digital signatures, randomized quantum handshake protocols, quantum APIs \\
        \bottomrule
    \end{tabularx}
\end{table*}

A layer-wise analysis is very important to fully understand the security of the quantum internet. Each layer has its own set of challenges. For example, the physical layer is very much limited by hardware flaws and environmental noise, while on the other hand, the link and network layers face issues with synchronization and trust. The application layer must unite quantum and classical security models for sure. One key comment is that many attacks work because current quantum networks still depend on classical communication channels. In the near future, as soon as the quantum systems improve, it will be a much more complex and sophisticated task to design flexible security measures that handle both quantum and classical weaknesses \citep{Preskill2018,Satoh2021}.

There are several open challenges however. Many current studies assume that network components are uniform, but real networks are not. Also, the security of routing protocols in an asynchronous environment is not yet fully proven, especially when attackers may exploit untrusted nodes \citep{Inamori2001}. Future work should develop dynamic security solutions that can adapt to network changes in real-time, much like what we see in the regular approaches.

One limitation worth mentioning here is that they do not fully integrate the fast-changing quantum states with the slower classical update processes. This delay in classical communication is a big problem because qubits lose their state quickly, and the framework does not completely solve this issue \citep{Preskill2018}. Also, while the framework provides a useful layer-by-layer breakdown, it sometimes assumes ideal conditions. In real networks, components can be very different, and link qualities may vary. Finally, the paper does not include extensive experimental tests or detailed simulations on irregular network topologies.

In simple words, the quantum internet is a groundbreaking technology that has its own set of unique features, however, because of them, it also creates a new set of security challenges that need a detailed, layer-by-layer analysis. This work sets out a security framework that covers vulnerabilities at different layers and suggests possible countermeasures. It is needless to say that, the framework shows promise, however, more research is needed to refine these strategies and also someone needs to test them in real-world networks (which will be the hardest part). Lastly, developing adaptive and strong security measures is important to ensure the long-term integrity and reliability of the quantum internet for practical use.

\section{Discussion}

This review has outlined two main paths forward in blockchain security under quantum threats: post-quantum cryptographic upgrades and fundamentally quantum-native systems. While both directions are promising, neither is without deep technical and practical challenges. This section identifies open problems and highlights research areas that demand further attention.

\subsection{Key Research Gaps}

Despite recent progress, significant gaps remain:
\begin{itemize}
    \item \textbf{Hardware Limitations:} Quantum blockchains require quantum memories, entangled state distribution, and error-corrected qubits—all of which remain experimental or unavailable at scale.
    \item \textbf{Routing Realism:} Most quantum routing models rely on idealized assumptions (e.g., perfect links, synchronized timing). Realistic simulations and testbeds are urgently needed.
    \item \textbf{Incentive Models:} Few papers explore economic or game-theoretic models for quantum consensus. The interaction between incentives, verification costs, and quantum randomness needs deeper exploration.
    \item \textbf{Cross-Layer Design:} Most current work focuses on individual protocol layers. Integrated approaches—considering quantum hardware constraints, security assumptions, and blockchain architecture—are rare.
    \item \textbf{Lack of Testbeds:} Experimental validation is minimal. Even hybrid quantum-classical blockchain prototypes are largely untested outside simulations.
\end{itemize}

\subsection{Design Trade-offs}
Post-quantum cryptographic schemes introduce computational and storage overhead, which may not be acceptable in low-latency or resource-constrained environments. Quantum blockchain designs, while theoretically secure, demand advanced hardware and suffer from limited throughput, poor scalability, and operational fragility. Researchers must explicitly weigh these trade-offs when designing protocols.

Future designs may benefit from hybrid architectures that combine:
\begin{itemize}
    \item Classical infrastructure with post-quantum cryptographic primitives.
    \item Quantum key distribution or quantum random number generators with classical consensus models.
    \item Modular architectures that can evolve alongside hardware capabilities.
\end{itemize}
Such systems can act as transition stages before fully quantum blockchains become viable.

Thereby, this work propose a phased adoption strategy:
\begin{enumerate}
    \item Transition classical blockchains to NIST-backed post-quantum cryptography.
    \item Build testbeds for hybrid quantum-classical blockchains using QKD modules.
    \item Invest in small-scale quantum consensus experiments using optical networks or simulators.
    \item Establish open benchmarks and simulation frameworks for routing, consensus, and scalability under quantum constraints.
\end{enumerate}

Quantum computing will eventually transform trust and security infrastructure. Whether via resilient mathematical schemes or quantum-native protocols, the blockchain community must prepare now. Building secure, scalable, and hardware-aware blockchain protocols will require not just new algorithms but also real-world validation, interdisciplinary collaboration, and significant investment in quantum infrastructure.

\section{Conclusion}

Quantum computing is no longer a distant possibility—it is an active and growing field with direct implications for digital trust systems. This paper has examined how classical blockchains, which rely on public-key cryptography and hash functions, are fundamentally vulnerable to quantum attacks. In response, two major directions have emerged: post-quantum blockchain designs that extend classical security with quantum-resistant cryptography, and quantum blockchain models that build security directly into the quantum layer using entanglement and QKD.

Through a review of foundational papers, this work has highlighted key technical proposals, common design patterns, and unresolved challenges in routing, consensus, scalability, and real-world deployment. The comparative analysis clarified the distinct trade-offs between post-quantum and quantum approaches, noting that while post-quantum solutions are more deployment-ready, quantum blockchains offer theoretically stronger guarantees but rely on hardware that is not yet scalable.

Moving forward, the field must address critical research gaps such as hardware-aware design, realistic simulation environments, cross-layer integration, and formal consensus models. Hybrid systems and phased deployments may serve as a bridge between classical and quantum-secure infrastructures.

In sum, preparing blockchain systems for the quantum era requires both cautious upgrades and bold experimentation. By identifying key vulnerabilities, surveying cutting-edge proposals, and outlining paths forward, this paper aims to support ongoing research that will shape the future of secure decentralized technologies.



\begin{thebibliography}{99}

\bibitem[Shor(1994)]{Shor} 
P. W. Shor, ``Algorithms for quantum computation: Discrete logarithms and factoring,'' in \emph{Proc. 35th Annu. Symp. Found. Comput. Sci.}, 1994, pp. 124--134, doi: 10.1109/SFCS.1994.365700.

\bibitem[Kearney(2021)]{Kearney} 
J. J. Kearney and C. A. Perez-Delgado, ``Vulnerability of blockchain technologies to quantum attacks,'' \emph{Array}, vol. 10, 2021, Art. no. 100065, doi: 10.1016/j.array.2021.100065.

\bibitem[RajanVisser(2019)]{RajanVisser} 
D. Rajan and M. Visser, ``Quantum blockchain using entanglement in time,'' \emph{Quantum Rep.}, vol. 1, no. 1, pp. 3--11, 2019, doi: 10.3390/quantum1010002.

\bibitem[Simon(2017)]{Simon} 
C. Simon, ``Towards a global quantum network,'' \emph{Nature Photon.}, vol. 11, pp. 678--680, 2017, doi: 10.1038/s41566-017-0032-0.

\bibitem[Zhang(2009)]{Zhang} 
X. Zhang, ``One-way quantum identity authentication based on public key,'' \emph{Chin. Sci. Bull.}, vol. 54, pp. 2018--2021, 2009, doi: 10.1007/s11434-009-0350-9.

\bibitem[SunSopek(2019)]{SunSopek} 
X. Sun, M. Sopek, Q. Wang, and P. Kulicki, ``Towards quantum-secured permissioned blockchain: Signature, consensus, and logic,'' \emph{Entropy}, vol. 21, no. 9, 2019, Art. no. 887, doi: 10.3390/e21090887.

\bibitem[Edwards(2020)]{Edwards} 
M. Edwards, A. Mashatan, and S. Ghose, ``A review of quantum and hybrid quantum/classical blockchain protocols,'' \emph{Quantum Inf. Process.}, vol. 19, 2020, Art. no. 184, doi: 10.1007/s11128-020-02672-y.

\bibitem[Kiktenko(2018)]{Kiktenko} 
E. O. Kiktenko et al., ``Quantum-secured blockchain,'' \emph{Quantum Sci. Technol.}, vol. 3, no. 3, 2018, Art. no. 035004, doi: 10.1088/2058-9565/aabc6b.

\bibitem[SunKulicki(2018)]{SunKulicki} 
X. Sun, Q. Wang, P. Kulicki, and X. Zhao, ``Quantum-enhanced logic-based blockchain I: Quantum honest-success Byzantine agreement and qulogicoin,'' 2018, arXiv:1805.06768, doi: 10.48550/arXiv.1805.06768.

\bibitem[SalmanEtAl2019(2019)]{SalmanEtAl2019} 
T. Salman, M. Zolanvari, A. Erbad, R. Jain, and M. Samaka, ``Security services using blockchains: A state of the art survey,'' \emph{IEEE Commun. Surv. Tut.}, vol. 21, no. 1, pp. 858--880, Jan.--Mar. 2019, doi: 10.1109/COMST.2018.2863956.

\bibitem[NoferEtAl2017(2017)]{NoferEtAl2017} 
M. Nofer, P. Gomber, O. Hinz, and D. Schiereck, ``Blockchain,'' \emph{Bus. Inf. Syst. Eng.}, vol. 59, no. 3, pp. 183--187, 2017, doi: 10.1007/s12599-017-0467-3.

\bibitem[BernsteinLange(2017)]{BernsteinLange} 
D. Bernstein and T. Lange, ``Post-quantum cryptography,'' \emph{Nature}, vol. 549, pp. 188--194, 2017, doi: 10.1038/nature23461.

\bibitem[FernandezCarames(2020)]{FernandezCarames} 
T. M. Fernández-Caramès and P. Fraga-Lamas, ``Towards post-quantum blockchain: A review on blockchain cryptography resistant to quantum computing attacks,'' \emph{IEEE Access}, vol. 8, pp. 21091--21116, 2020, doi: 10.1109/ACCESS.2020.2968985.

\bibitem[Greenberger(1989)]{Greenberger} 
D. Greenberger, M. A. Horne, and A. Zeilinger, ``Going beyond bell’s theorem,'' in \emph{Bell’s Theorem, Quantum Theory and Concept. of the Universe (Fundamental Theories of Physics)}, vol. 37. Dordrecht, The Netherlands: Springer, 1989, \url{doi: 10.1007/978-94-017-0849-4_10}

\bibitem[Carvacho(2017)]{Carvacho} 
G. Carvacho, F. Graffitti, V. D’Ambrosio, B. C. Hiesmayr, and F. Sciarrino, ``Experimental investigation on the geometry of GHZ states,'' \emph{Sci. Rep.}, vol. 7, 2017, Art. no. 13265, doi: 10.1038/s41598-017-13124-6.

\bibitem[Megidish(2013)]{Megidish} 
E. Megidish, A. Halevy, T. Shacham, T. Dvir, L. Dovrat, and H. S. Eisenberg, ``Entanglement between photons that have never coexisted,'' \emph{Phys. Rev. Lett.}, vol. 110, 2013, Art. no. 210403, doi: 10.1103/PhysRevLett.110.210403.

\bibitem[McCutcheon(2016)]{McCutcheon} 
W. McCutcheon et al., ``Experimental verification of multipartite entanglement in quantum networks,'' \emph{Nature Commun.}, vol. 7, 2016, Art. no. 13251, doi: 10.1038/ncomms13251.

\bibitem[EthereumWhitepaper(2014)]{EthereumWhitepaper} 
``Ethereum whitepaper,'' Ethereum.org, Accessed: Jun. 14. 2014. [Online]. Available: https://ethereum.org/en/whitepaper/

\bibitem[TannuQureshi2019(2019)]{TannuQureshi2019} 
S. S. Tannu and M. K. Qureshi, ``Not all qubits are created equal: A case for variability-aware policies for NISQ-Era quantum computers,'' in \emph{Proc. 24th Int. Conf. Architectural Support Program. Lang. Oper. Syst.}, 2019, pp. 987--999, doi: 10.1145/3297858.3304007.

\bibitem[RodenburgPappas2017(2017)]{RodenburgPappas2017} 
B. Rodenburg and S. P. Pappas, ``Blockchain and quantum computing,'' MITRE Corp., Bedford, MA, USA, Tech. Rep., 2017. Accessed: Jun. 14. 2022. [Online]. Available: https://apps.dtic.mil/sti/citations/AD1125436

\bibitem[Bruschi(2014)]{Bruschi} 
D. E. Bruschi, C. Sabín, A. White, V. Baccetti, D. K. L. Oi, and I. Fuentes, ``Testing the effects of gravity and motion on quantum entanglement in space-based experiments,'' \emph{New J. Phys.}, vol. 16, 2014, Art. no. 053041, doi: 10.1088/1367-2630/16/5/053041.

\bibitem[Banerjee(2020)]{Banerjee} 
S. Banerjee, A. Mukherjee, and P. K. Panigrahi, ``Quantum blockchain using weighted hypergraph states,'' \emph{Phys. Rev. Res.}, vol. 2, 2020, Art. no. 013322, doi: 10.1103/PhysRevResearch.2.013322.

\bibitem[Nowakowski(2017)]{Nowakowski} 
M. Nowakowski, ``Quantum entanglement in time,'' in \emph{Proc. AIP Conf. Proc.}, 2017, Art. no. 020007, doi: 10.1063/1.4982771.

\bibitem[RajanArxiv(2020)]{RajanArxiv} 
D. Rajan, ``Quantum entanglement in time,'' 2020, arXiv:2007.05969, doi: 10.48550/arXiv.2007.05969.

\bibitem{YangSurvey}
Z. Yang, M. Zolanvari, and R. Jain, ``A Survey of Important Issues in Quantum Computing and Communications,'' \emph{IEEE Communications Surveys \& Tutorials}, vol.~25, no.~2, 2023.

\bibitem{Chow}
J. Chow, O. Dial, and J. Gambetta, ``IBM Quantum Breaks the 100-Qubit Processor Barrier,'' 2021.

\bibitem{Roberts}
S. Roberts, ``This New Startup Has Built a Record-Breaking 256-Qubit Quantum Computer,'' 2021.

\bibitem{Reilly}
D. J. Reilly, ``Challenges in scaling-up the control interface of a quantum computer,'' in \emph{Proc. IEEE Int. Electron Devices Meeting (IEDM)}, 2019, pp.~31.7.1--31.7.6.

\bibitem{XuQKD}
F. Xu, X. Ma, Q. Zhang, H. Lo, and J. Pan, ``Secure quantum key distribution with realistic devices,'' \emph{Rev. Mod. Phys.}, vol.~92, no.~2, 2020.

\bibitem{PerdomoOrtiz}
A. Perdomo-Ortiz, M. Benedetti, J. Realpe-G\'omez, and R. Biswas, ``Opportunities and challenges for quantum-assisted machine learning in near-term quantum computers,'' \emph{Quantum Science and Technology}, vol.~3, no.~3, 2018.

\bibitem{Sangouard}
N. Sangouard, C. Simon, H. De Riedmatten, and N. Gisin, ``Quantum repeaters based on atomic ensembles and linear optics,'' \emph{Rev. Mod. Phys.}, vol.~83, 2011.

\bibitem{ChenSpace}
Y. Chen \emph{et al.}, ``An integrated space-to-ground quantum communication network over 4,600 kilometers,'' \emph{Nature}, vol.~589, 2021.

\bibitem{Mosca}
M. Mosca, ``Cybersecurity in an era with quantum computers: will we be ready?'' \emph{IEEE Security \& Privacy}, vol.~16, no.~5, 2018.

\bibitem{Biamonte}
J. Biamonte, P. Wittek, N. Pancotti, P. Rebentrost, N. Wiebe, and S. Lloyd, ``Quantum machine learning,'' \emph{Nature}, vol.~549, 2017.

\bibitem{BennettBrassard}
C. H. Bennett and G. Brassard, ``Quantum cryptography: public key distribution and coin tossing,'' in \emph{Proc. IEEE Int. Conf. on Computers, Systems, and Signal Processing}, 1984, pp.~175--179.

\bibitem{BroadbentSchaffner}
A. Broadbent and C. Schaffner, ``Quantum cryptography beyond quantum key distribution,'' \emph{Designs, Codes and Cryptography}, vol.~78, 2015.


\bibitem[Nakamoto(2008)]{Nakamoto2008} Nakamoto, S., \emph{Bitcoin: A Peer-to-Peer Electronic Cash System}, 2008.
\bibitem[Johnson et al.(2001)]{Johnson2001} Johnson, D., Menezes, A., \& Vanstone, S., \emph{The Elliptic Curve Digital Signature Algorithm (ECDSA)}, International Journal of Information Security, 2001.

\bibitem[Grover(1996)]{Grover1996} Grover, L. K., \emph{A Fast Quantum Mechanical Algorithm for Database Search}, Proc. 28th Annual ACM Symposium on Theory of Computing, 1996.
\bibitem[Bennett \& Brassard(2014)]{Bennett2014} Bennett, C. H. \& Brassard, G., \emph{Quantum Cryptography: Public Key Distribution and Coin Tossing}, Theoretical Computer Science, 2014.

\bibitem[Ekert(1991)]{Ekert1991} Ekert, A. K., \emph{Quantum Cryptography Based on Bell’s Theorem}, Physical Review Letters, 1991.

\bibitem[Ajtai(1996)]{Ajtai1996} Ajtai, M., \emph{Generating Hard Instances of Lattice Problems}, Proc. 28th Annual ACM Symposium on Theory of Computing, 1996.
\bibitem[Regev(2009)]{Regev2009} Regev, O., \emph{On Lattices, Learning with Errors, Random Linear Codes, and Cryptography}, Journal of the ACM, 2009.
\bibitem[Kiktenko et al.(2018)]{Kiktenko2018} Kiktenko, E. O., et al., \emph{Quantum-Secured Blockchain}, Quantum Science and Technology, 2018.
\bibitem[Rajan \& Visser(2019)]{Rajan2019} Rajan, D. \& Visser, M., \emph{Quantum Blockchain Using Entanglement in Time}, Quantum Reports, 2019.
\bibitem[JPMorgan(2018)]{JPMorgan2018} JPMorgan, \emph{Research on a Quantum-Resistant Blockchain Network}, Cointelegraph, 2018.


\bibitem[Kimble(2008)]{Kimble2008}
H.~J. Kimble, ``The quantum internet,'' \emph{Nature}, vol. 453, pp. 1023, 2008.

\bibitem[Gisin and Thew(2007)]{GisinThew2007}
N.~Gisin and R.~Thew, ``Quantum communication,'' \emph{Nature Photonics}, vol. 1, pp. 165--171, 2007.

\bibitem[Komar et al.(2014)]{Komar2014}
P.~Komar \emph{et al.}, ``A quantum network of clocks,'' \emph{Nature Physics}, vol. 10, pp. 582--587, 2014.

\bibitem[Pant et al.(2019)]{Pant2019}
M.~Pant, H.~Krovi, D.~Towsley, L.~Tassiulas, L.~Jiang, P.~Basu, D.~Englund, and S.~Guha, ``Quantum routing for the quantum internet,'' \emph{npj Quantum Inf.}, vol. 5, p. 25, 2019.

\bibitem[Chakraborty et al.(2019)]{Chakraborty2019}
K.~Chakraborty, F.~Rozpedek, A.~Dahlberg, and S.~Wehner, ``Distributed routing in a quantum internet,'' arXiv:1907.11630, 2019.

\bibitem[Briegel et al.(1998)]{Briegel1998}
H.-J. Briegel, W.~Dür, J.~I. Cirac, and P.~Zoller, ``Quantum repeaters: The role of imperfect local operations in quantum communication,'' \emph{Phys. Rev. Lett.}, vol. 81, pp. 5932--5935, 1998.

\bibitem[Pan et al.(1998)]{Pan1998}
J.-W. Pan, D.~Bouwmeester, H.~Weinfurter, and A.~Zeilinger, ``Experimental entanglement swapping: Entangling photons that never interacted,'' \emph{Phys. Rev. Lett.}, vol. 80, pp. 3891--3894, 1998.

\bibitem[Tchebotareva et al.(2019)]{Tchebotareva2019}
A.~Tchebotareva, S.~L.N. Hermans, P.~C. Humphreys, D.~Voigt, P.~J. Harmsma, L.~K. Cheng, A.~L. Verlaan, N.~Dijkhuizen, W.~de Jong, \emph{et al.}, ``Extending the reach of quantum communications using telecom wavelengths,'' \emph{Phys. Rev. Lett.}, vol. 123, p. 063601, 2019.

\bibitem[Alexander et al.(2012)]{Alexander2012}
R.~Alexander, A.~Brandt, J.~P. Vasseur, J.~Hui, K.~Pister, P.~Thubert, P.~Levis, R.~Struik, R.~Kelsey, \emph{et al.}, ``RPL: IPv6 Routing Protocol for Low-Power and Lossy Networks,'' RFC 6550, 2012.

\bibitem[Gallager et al.(1983)]{Gallager1983}
R.~G. Gallager, P.~A. Humblet, and P.~M. Spira, ``A distributed algorithm for minimum spanning trees,'' \emph{ACM Trans. Program. Lang. Syst.}, vol. 5, pp. 66--77, 1983.

\bibitem[Degen et al.(2017)]{DegenReinhardCappellaro2017}
C.~L. Degen, F.~Reinhard, and P.~Cappellaro, ``Quantum sensing,'' \emph{Rev. Mod. Phys.}, vol. 89, p. 035002, 2017.

\bibitem[Beals et al.(2013)]{Beals2013}
R.~Beals, S.~Brierley, O.~Gray, A.~W. Harrow, S.~Kutin, N.~Linden, D.~Shepherd, and M.~Stather, ``Efficient distributed quantum computing,'' \emph{Proc. R. Soc. A}, vol. 469, p. 20120686, 2013.

\bibitem[Van Meter and Devitt(2016)]{VanMeterDevitt2016}
R.~Van Meter and S.~J. Devitt, ``The path to scalable distributed quantum computing,'' \emph{Computer}, vol. 49, pp. 31--42, 2016.

\bibitem[Azuma et al.(2015)]{AzumaTamakiLo2015}
K.~Azuma, K.~Tamaki, and H.-K. Lo, ``All-photonic quantum repeaters,'' \emph{Nature Communications}, vol. 6, p. 6787, 2015.

\bibitem[Wehner et al.(2018)]{Wehner2018}
S.~Wehner, D.~Elkouss, and R.~Hanson, ``Quantum Internet: A Vision for the Road Ahead,'' \emph{Science}, vol. 362, no. 6412, p. eaam9288, 2018.

\bibitem[Yang et al.(2023)]{Yang2023}
Z.~Yang, M.~Zolanvari, and R.~Jain, ``A Survey of Important Issues in Quantum Computing and Communications,'' \emph{IEEE Communications Surveys \& Tutorials}, vol. 25, no. 2, pp. 1059--1094, 2023.

\bibitem[Lo et al.(2014)]{LoCurtyTamaki2014}
H.-K. Lo, M.~Curty, and K.~Tamaki, ``Secure Quantum Key Distribution,'' \emph{Nature Photonics}, vol. 8, pp. 595--604, 2014.

\bibitem[Muralidharan et al.(2016)]{Muralidharan2016}
S.~Muralidharan \emph{et al.}, ``Optimal Architectures for Long Distance Quantum Communication,'' \emph{Scientific Reports}, vol. 6, p. 20463, 2016.

\bibitem[Shi and Qian(2020)]{ShiQian2020}
S.~Shi and C.~Qian, ``Concurrent Entanglement Routing for Quantum Networks: Model and Designs,'' in \emph{Proc. SIGCOMM '20}, New York, NY, USA: ACM, July 2020, pp. 62--75.

\bibitem[Yang et al.(2024)]{Yang2024}
Z.~Yang \emph{et al.}, ``Asynchronous Entanglement Routing for the Quantum Internet,'' \emph{AVS Quantum Sci.}, vol. 6, no. 1, Jan. 2024.

\bibitem[Furnas and Zacks(1994)]{FurnasZacks1994}
G.~W. Furnas and J.~Zacks, ``Multitrees: Enriching and Reusing Hierarchical Structure,'' in \emph{Proc. SIGCHI Conf.}, Boston, MA, USA: ACM, 1994, pp. 330--336.

\bibitem[Erds and Rnyi(1959)]{ErdsRnyi1959}
P.~Erds and A.~Rnyi, ``On Random Graphs,'' \emph{Publ. Math. Debrecen}, vol. 6, pp. 290--297, 1959.

\bibitem[Preskill(2018)]{Preskill2018}
J.~Preskill, ``Quantum Computing in the NISQ Era and Beyond,'' \emph{Quantum}, vol. 2, p. 79, 2018.

\bibitem[ESnet(2023)]{ESnet2023}
``About ESnet,'' ESnet; available: \url{https://www.es.net/about/}, accessed: Nov. 15, 2023.

\bibitem[Knight et al.(2011)]{Knight2011}
S.~Knight \emph{et al.}, ``The Internet Topology Zoo,'' \emph{IEEE J. Sel. Areas Commun.}, vol. 29, no. 9, pp. 1765--1775, Oct. 2011.

\bibitem[Internet2(2023)]{Internet2}
``About Internet2,'' Internet2; available: \url{https://internet2.edu/community/about-us/}, accessed: Nov. 15, 2023.


\bibitem[Brassard et al.(2000)]{Brassard2000}
G. Brassard, N. Lütkenhaus, T. Mor, and B. C. Sanders, ``Limitations on practical quantum cryptography,'' \emph{Phys. Rev. Lett.}, vol. 85, pp. 1330--1333, 2000.

\bibitem[Zhao et al.(2008)]{Zhao2008}
Y. Zhao, C. H. F. Fung, B. Qi, C. Chen, and H. K. Lo, ``Quantum hacking: Experimental demonstration of time-shift attack against practical quantum-key-distribution systems,'' \emph{Phys. Rev. A}, vol. 78, p. 042333, 2008.

\bibitem[Xu et al.(2005)]{Xu2005}
X. Ma, B. Qi, Y. Zhao, and H. K. Lo, ``Practical decoy state for quantum key distribution,'' \emph{Phys. Rev. A}, vol. 72, p. 012326, 2005.

\bibitem[Hwang(2003)]{Hwang2003}
W. Y. Hwang, ``Quantum Key Distribution with High Loss: Toward Global Secure Communication,'' \emph{Phys. Rev. Lett.}, vol. 91, p. 057901, 2003.

\bibitem[Satoh et al.(2021)]{Satoh2021}
T. Satoh et al., ``Attacking the Quantum Internet,'' \emph{IEEE Trans. Quantum Eng.}, vol. 2, pp. 1--17, 2021.

\bibitem[Inamori(2001)]{Inamori2001}
H. Inamori, ``Security of EPR-based quantum cryptography against incoherent symmetric attacks,'' \emph{J. Phys. A: Math. Gen.}, vol. 34, p. 6913, 2001.

\bibitem[Illiano et al.(2022)]{Illiano2022}
J. Illiano, M. Caleffi, A. Manzalini, and A. S. Cacciapuoti, ``Quantum Internet protocol stack: A comprehensive survey,'' \emph{Comput. Netw.}, vol. 213, p. 109092, 2022.

\bibitem[Lu et al.(2022)]{Lu2022}
C.-Y. Lu, Y. Cao, C.-Z. Peng, and J.-W. Pan, ``Micius quantum experiments in space,'' \emph{Rev. Mod. Phys.}, vol. 94, p. 035001, 2022.


\end{thebibliography}
\end{document}